\documentclass[final,a4paper,10pt]{article}
\usepackage[utf8]{inputenc}
\usepackage{epsfig}
\usepackage{varioref}
\usepackage{pstricks, pst-node, pst-grad, pst-plot}
\usepackage{amsmath}
\usepackage{amssymb}
\usepackage{latexsym}
\usepackage{mathrsfs}
\usepackage{xspace}
\usepackage{subfigure}
\usepackage{float}
\usepackage{multirow}
\usepackage{algorithm}
\usepackage{algpseudocode}
\usepackage{fancyhdr}
\usepackage[margin=.8cm,small,bf]{caption}
\usepackage{setspace} 
\usepackage{array}
\usepackage{listings}
\usepackage{pstricks-add} 
\usepackage[square,numbers,sort&compress]{natbib}

\floatstyle{plain}
\restylefloat{figure}
  {\refstepcounter{lemma}%
    \par\bigskip\noindent\textbf{Example \arabic{section}.\arabic{lemma}} }%
  {\hspace*{\stretch{1}}$\Diamond$\vspace{4mm}\par\noindent}

\newcommand{\mcS}{\mathcal{S}}

\newcommand{\mcA}{\mathcal{A}}
\newcommand{\mcT}{\mathcal{T}}

\title{Vertical partitioning of relational OLTP databases using integer programming}
\author{Rasmus Resen Amossen\footnote{email: rasmus@resen.org}\\IT University of Copenhagen}
\date{January 2010}
\begin{document}
\maketitle
\begin{abstract}
A way to optimize performance of relational row store databases is to reduce the row widths by vertically partitioning tables into table fractions in order to minimize the number of irrelevant columns/attributes read by each transaction.
This paper considers vertical partitioning algorithms for relational row-store OLTP databases with an H-store-like architecture, meaning that we would like to maximize the number of single-sited transactions.
We present a model for the vertical partitioning problem that, given a schema together with a vertical partitioning and a workload, estimates the costs (bytes read/written by storage layer access methods and bytes transferred between sites) of evaluating the workload on the given partitioning.
The cost model allows for arbitrarily prioritizing load balancing of sites vs. total cost minimization.
We show that finding a minimum-cost vertical partitioning in this model is NP-hard and present two algorithms returning solutions in which single-sitedness of read queries is preserved while allowing column replication (which may allow a drastically reduced cost compared to disjoint partitioning).
The first algorithm is a quadratic integer program that finds optimal minimum-cost solutions with respect to the model, and the second algorithm is a more scalable heuristic based on simulated annealing.
Experiments show that the algorithms can reduce the cost of the model objective by 37\% when applied to the TPC-C benchmark and the heuristic is shown to obtain solutions with cost close to the ones found using the quadratic program.
\end{abstract}

\section{Introduction}
In this paper we consider OLTP databases with an H-store \cite{hstore} like architecture in which we would aim for maximizing the number of single-sited transactions (i.e. transactions that can be run to completion on a single site).
Given a database schema and a workload we would like to reduce the cost of evaluating the workload.
In row-stores, where each row is stored as a contiguous segment and access is done in quantums of whole rows, a significant amount of superfluous columns/attributes (we will use the term \emph{attribute} in the following) are likely to be accessed during evaluation of a workload.
It is easy to see that this superfluous data access may have a negative impact on performance so in an optimal world the amount of data accessed by each query should be minimized.
One approach to this is to perform a \emph{vertical partitioning} of the tables in the schema.
A vertical partitioning is a, possibly non-disjoint, distribution of attributes and transactions onto multiple physical or logical sites.
(Notice, that vertical and horizontal partitioning are not mutually exclusive and can perfectly be used together).
The optimality of a vertical partitioning depends on the context:
OLAP applications with lots of many-row aggregates will likely benefit from parallelizing the transactions on multiple sites and exchanging small sub-results between the sites after the aggregations.
OLTP applications on the other hand, with many short-lived transactions, no many-row aggregates and with few or no few-row aggregates would likely benefit from gathering all attributes read by a query locally on the same site: inter-site transfers and the synchronization mechanisms needed for non-single-sited or parallel queries (e.g. undo and redo logs) are assumed to be bottlenecks in situations with short transaction durations.
\citet{hstore} and \citet{kallman08} discuss the benefits of single-sitedness in high-throughput OLTP databases in more details.

This paper presents a cost model together with two algorithms that find either optimal or close-to-optimal vertical partitionings with respect to the cost model.
The two algorithms are based on quadratic programming and simulated annealing, respectively.
For a given partitioning and a workload, the cost model estimates the number of bytes read/written by access methods in the storage layer and the amount of data transfer between sites.
Our model is made with a specific setting in mind, captured by five headlines:
\begin{description}
\item[OLTP] The database is a transaction processing system with many short lived transactions.
\item[Aggregates] No many-row aggregates and few (or no) aggregates on small row-subsets.
\item[Preserve single-sitedness] We should try to avoid breaking single-sitedness as a large number of single-sited transactions will reduce the need for inter-site transfers and completely eliminate the need for undo and redo logs for these queries if the partitioning is performed on an H-store like DMBS \cite{hstore}.
\item[Workload known] Transactions used in the workload together with some run-time statistics are assumed to be known when applying the algorithms.
\end{description}
Furthermore, following the consensus in the related work (see Section \ref{sec:related_work}) we simplify the model by not considering time spent on network latency (if all vertical partitions are placed locally on a single site, then time spend on network latency is trivially zero anyway).
A description of how to include latency in the model at the expense of increased complexity can be found in Appendix~\ref{sec:latency}.

\subsection{Outline of approach}
The basic idea is as follows.
We are given an input in form of a schema together with a workload in which queries are grouped into transactions, and each query is described by a set of statistical properties.

For each query $q$ in the workload and for each table $r$ accessed by $q$ the input provides the average number $n_r$ of rows from table $r$ that is retrieved from or written to storage by query $q$.
Together with the (average) width $w_a$ of each attribute $a$ from table $r$ this generally gives a good estimate for how much attribute $a$ costs in retrievals/writes by access methods for each evaluation of query $q$, namely $W'_{a,q} = w_a\cdot n_r$.

Given a set of sites, the challenge is now to find a non-disjoint distribution of all attributes, and a disjoint distribution of transactions to these sites so that the costs of retrievals, writes and inter-site transfers, each defined in terms of $W'_{a,q}$ as explained in details below, is minimized.
This means, that the primary executing site of any given query is assumed to be the site that hosts the transaction holding that query.

As mentioned above, our algorithms will not break single-sitedness for read queries and therefore no additional costs are added to the execution of read queries by applying this algorithm.
In contrast, since the storage costs (the sum of retrieval, write and inter-site transfer costs) for a query is minimized and each tuple become as narrow as possible, the total costs of evaluating the queries (e.g. processing joins, handling intermediate subresults, etc.) are assumed to be, if not minimized, then reduced too.


\subsection{Contributions}
This paper contributes with the following:
\begin{itemize}
\item an algorithm optimized for H-store like architectures, preserving single-sitedness for read queries and in which load balance among sites versus minimization of total costs can be prioritized arbitrarily,
\item a more scalable heuristic, and
\item a micro benchmark of a) both algorithms based on TPC-C and a set of random instances, b) a comparison between the benefits of local versus remote partition location, and c) a comparison between disjoint and non-disjoint partitioning.
\end{itemize}

\subsection{Related work}\label{sec:related_work}
A lot of work has been done on data allocation and vertical partitioning but to the best of our knowledge, no work solves the exact same problem as the present paper: distributing both transactions and attributes to a set of sites, allowing attribute replication, preserving single-sitedness for read queries and prioritizing load balancing vs. total cost minimization.
We therefore order the references below by increasing estimated problem similarity and do not mention work dedicated on vertical partitioning of OLAP databases.

In \citeyear{eisner76} \citet{eisner76} reduced the cost of information retrieval by vertically partitioning records into a primary and a secondary record segment.
This was done by constructing a bi-partite graph with two node sets: one set with a node for each attribute and one set with a node for each transaction.
By connecting attribute and transaction nodes with a weighted edge according to their affinity, a min-cut algorithm could be applied to construct the partitioning.

\citet{sacca85} assumed a set of horizontal and vertical fragments of a database was known in advance and produced a disjoint distribution of these fragments onto a set of network-connected processors using a greedy first-fit bin packing heuristic.
Similarly, \citet{menon05} distributed a set of predefined fragments to a set of sites, but used a linearized quadratic program to compute the solution.

\citet{vpart_sarathy97} took as input a geographically distributed database together with statistics for a query pattern on this database and produced as output a non-disjoint distribution of whole database tables to the physical sites so that the total amount of transfer was minimized.
They modelled the problem as a linearized quadratic program which was solved in practice using heuristics.
The costs of joins were minimized by first transferring join keys and then transferring the relevant attributes for the relevant rows to a single collector site.

\citet{vpart_navathe89} constructed a disjoint partitioning with non-remote partition placement.
They used an attribute affinity matrix to represent a complete weighted graph and generated partitions by finding a linearly connected spanning tree and considering a cycle as a fragment.

\citet{vpart_cornel90} generated a non-remote, disjoint partitioning minimizing the amount of disk access by recursively applying a binary partitioning.
The partitioning decisions were based on an integer program and with strong assumptions on a System-R like architecture when estimating the amount of disk access.

\citet{agrawal04} also constructed a disjoint partitioning with non-remote partition placement.
They used a two-phase strategy where the first phase generated all relevant attribute groups using association rules \cite{agarwal01} considering only one query at a time, and the second phase merged the attribute groups that were useful across queries.

\citet{son04} presented an algorithm for generating disjoint partitioning by either minimizing costs or by ensuring that exactly $k$ vertical fragments were produced.
Inter-site transfer costs were not considered.
The partitioning was produced using a bottom-up strategy, iteratively merging two selected partitions with the best ``merge profit'' until only one large super-partition existed.
The $k$-way partitioning was found at the iteration having exactly $k$ partitions and the lowest-cost partitioning was found at the iteration with the lowest cost.

\citet{chu93} minimized the amount of disk access by constructing a non-remote and non-disjoint vertical partitioning.
Two binary partitioning algorithms based on the branch-and-bound method were presented with varying complexity and accuracy.
The partitionings were formed by recursively applying the binary partitioning algorithms on the set of ``reasonable cuts''.

\citet{chakravarthy94} did not present an algorithm but gave an interesting objective function for evaluating vertical partitionings.
The function was based on the square-error criterion as given in \cite{jain88} for data clustering, but did not cover placement of transactions which, in our case, has a large influence on the expected costs.

\citet{vpart_navathe84} considered the vertical partitioning problem for three different environments: a) single site with one memory level, b) single site with several memory levels, and c) multiple sites.
The partitions could be both disjoint and non-disjoint.
A clustering algorithm grouped attributes with high affinity by using an attribute affinity matrix together with a bond energy algorithm \cite{mccormickjr72}.
Three basic algorithms for generating partitions were presented which, depending on the desired environment, used different prioritization of four access and transfer cost classes.


\subsection{Outline of paper}
In section \ref{sec:model} we derive a cost model together with a quadratic program defining the first algorithm.
Section \ref{sec:sa} describes a heuristic based on the cost model found in Section \ref{sec:model}, and Section \ref{sec:improvements} discusses a couple of ideas for improvements.
Computational results are shown in Section \ref{sec:results}.

\section{A linearized QP approach}\label{sec:model}
In this section we develop our base model, a quadratic program (QP), which will later be extended to handle load balancing and then linearized in order to solve it using a conventional mixed integer program (MIP) solver.

\subsection{The base model}
In a vertical partitioning for a schema and a workload we would like to minimize the sum
\begin{equation}\label{eq:a+pb}
A+pB
\end{equation}
where $A$ is the amount of data accessed locally in the storage layer, $B$ is the amount of data needed to be transferred over the network during query updates and $p$ is a penalty factor.

We assume that each transaction has a primary executing site.
For each transaction $t\in \mcT$, each table attribute $a \in \mcA$, and each site $s\in \mcS$ consider two decision variables $x_{t,s} \in \{0,1\}$ and $y_{a,s} \in \{0,1\}$ indicating if transaction $t$ is executed on site $s$ and if attribute $a$ is located on site $s$, respectively.
All transactions must be located at exactly one site (their primary executing site), that is
\begin{equation}\label{eq:cons_x}
\sum_{t\in \mcT} x_{t,s} = 1\quad ,\forall s\in S
\end{equation}
and all attributes must be located at at least one site, that is
$$
\sum_{a\in \mcA} y_{a,s} \geq 1\quad ,\forall s\in S.
$$

To determine the size of $A$ and $B$ from equation \eqref{eq:a+pb} introduce five new static binary constants describing the database schema:
\begin{itemize}
\item $\alpha_{a,q}$ indicates if attribute $a$ itself is accessed by query $q$
\item $\beta_{a,q}$ indicates if attribute $a$ is part of a table that $q$ accesses
\item $\gamma_{q,t}$ indicates if query $q$ is used in transaction $t$
\item $\delta_q$ indicates if query $q$ is a write query
\item $\varphi_{a,t}$ indicates if any query in transaction $t$ reads attribute $a$
\end{itemize}
Single-sitedness should be maintained for reads.
That is, if a read query in transaction $t$ accesses attribute $a$ then $a$ and $t$ must be co-located:
$$
x_{t,s} \varphi_{a,t} = 1 \Rightarrow y_{a,s} = 1\quad ,\forall t\in \mcT, a\in \mcA
$$
or equivalently
$$
y_{a,s} - x_{t,s}  \varphi_{a,t} \geq 0 \quad ,\forall t\in \mcT, a\in \mcA.
$$

In order to estimate the cost of reading, writing and transferring data, introduce the following weights:
\begin{itemize}
\item $w_a$ denotes the average width of attribute $a$
\item $f_q$ denotes the frequency of query $q$
\item $n_{a,q}$ denotes for query $q$ the average number of rows retrieved from or written to the table holding attribute $a$
\end{itemize}
Then the cost of reading or writing $a$ in query $q$ is estimated to $W_{a,q} = w_a\cdot f_q\cdot n_{a,q}$ and the cost of transferring attribute $a$ over the network is estimated to $pW_{a,q}$.
Notice, that $W_{a,q}$ is only an estimate due to $f_q$ and $n_{a,q}$.

Consider the amount of local data access, $A$, and let $A = A_\text{R} + A_{W}$ where $A_\text{R}$ and $A_\text{W}$ is the amount of read and write access, respectively.
For a given site $r$ and query $q$, $A_\text{R}$ is the sum of all attribute weights $W_{a,q}$ for which 1)~$q$ is a read query, 2)~attribute $a$ is stored on $r$, 3)~the transaction that executes query $q$ is executed on $r$ and 4)~$q$ accesses any attribute in the table fraction that holds $a$.
As we maintain single-sitedness for reads, $\beta_{a,q}$ can be used to handle 4), resulting in
$$
A_\text{R} = \sum_{a,t,s,q} W_{a,q} \beta_{a,q} \gamma_{q,t} (1-\delta_q) x_{t,s} y_{a,s}.
$$
Accounting for local access of write queries, $A_\text{W}$, is less trivial. Consider the following three approaches:
\begin{description}
\item[Access relevant attributes] An attribute $a$ at site $s$ should be accounted for if and only if there exists an attribute $a'$ on $s$ that $q$ updates so that $a$ and $a'$ are attributes of the same table.
While this accounting is the most accurate of the three it is also the most expensive as it implies an element of the form $y_{a,s}y_{a',s}$ in the objective function which adds an undesirable amount of $|\mcA|^2|\mcS|$ variables and $3|\mcA|^2|\mcS|$ constraints to the problem when linearized (see Section~\ref{sec:linearization}).
\item[Access all attributes] We can get around the increased complexity by assuming that write queries $q$ always writes to all sites containing table fractions of tables accessed by $q$, regardless of whether $q$ actually accesses any of the attributes of the fractions.
While this is correct for insert statements (assuming that inserts always write complete rows) it is likely an overestimation for updates: imagine a lot of single-attribute updates on a wide table where the above method would have split the attribute in question to a separate partition.
This overestimation will imply that the model will partition tables that are updated often or replicate attributes less often than the accounting model described above.
\item[Access no attributes] Another approach to simplify the cost function is to completely avoid accounting for local access for writes and solely let the network transfer define the write costs.
With this underestimation of write costs, attributes will then tend to be replicated more often than in the first accounting model.
\end{description}
In this paper we choose the second approach, which gives a conservative overestimate of the write costs as we then obtain more accurate costs for inserts and avoid extending the model with undesirably many variables and constraints.
Intuitively speaking, this choice implies that read queries will tend to partition the tables for best possible read-performance, and the write queries will tend to minimize the amount of attribute replication.
We now have
$$
A_\text{W} = \sum_{q,a,s}W_{a,q}\beta_{a,q}\delta_q y_{a,s}
$$
and thus
\begin{equation}\label{eq:a}
A = \sum_{a,t,s,q} W_{a,q} \beta_{a,q} \gamma_{q,t} (1-\delta_q) x_{t,s} y_{a,s} + \sum_{q,a,s}W_{a,q}\beta_{a,q}\delta_q y_{a,s}.
\end{equation}

$B$ accounts for the amount of network transfer and since we enforce single-sitedness for all reads $B$ is solely the sum of transfer costs for write queries.
We assume that write queries only transfer the attributes they update and does not transfer to the site that holds their own transaction:
\begin{align*}
B & = \sum_{a,t,s,q}W_{a,q} \alpha_{a,q} \gamma_{q,t} \delta_q  (1-x_{t,s})  y_{a,s}. \label{eq:b}
\end{align*}
By noticing that $\sum_{a,t,s,q}\alpha_{a,q}\gamma_{q,t}y_{a,s} = \sum_{a,s,q}\alpha_{a,q}y_{a,s}$ we can construct the minimization problem as
\begin{equation}\label{eq:noloadbalance}
\begin{array}{lrll}
\min \quad       & \multicolumn{3}{l}{\sum_{t,a,s}c_1(a,t) x_{t,s} y_{a,s} + \sum_{a,s}c_2(a)  y_{a,s}}\\
\text{s.t.}\quad & \sum_s x_{t,s} &= 1     & \forall t\\
            & \sum_s y_{a,s}      &\geq 1       & \forall a\\
            & y_{a,s}-x_{t,s}\varphi_{a,t} &\geq 0 & \forall a, t\\
            & x_{t,s}, y_{a,s} &\in \{0,1\} & \forall t, a, s
\end{array}
\end{equation}
where
$$
c_1(a,t) = \sum_{q} W_{q,a} \gamma_{q,t} (\beta_{a,q}(1-\delta_q) - p \alpha_{a,q} \delta_q)
$$
and
$$
c_2(a) = \sum_{q} W_{a,q}\delta_q(\beta_{a,q}+p\alpha_{a,q}).
$$
Both $c_1$ and $c_2$ are completely induced by the schema, query workload and statistics and can therefore be considered static when the partitioning process starts.

\subsection{Adding load balancing}
We are interested in extending the model in \eqref{eq:noloadbalance} to also handle load balancing of the sites instead of just minimizing the sum of all data access/transfer.
From equation \eqref{eq:a} define the work of a single site $s\in \mcS$ as
\begin{equation}\label{eq:a_r}
\sum_{a,t} c_3(a,t)x_{t,s} y_{a,s} + \sum_{a}c_4(a)y_{a,s}
\end{equation}
where $c_3(a,t) = \sum_q W_{a,q}\gamma_{q,t}\beta_{a,q}(1-\delta_q)$ and $c_4(a) = \sum_q W_{a,q}\beta_{a,q}\delta_q$.
Introduce the variable $m$ and for each site $s$ let the value of \eqref{eq:a_r} be a lower bound for $m$.
Adding $m$ to the objective function is then equivalent to also minimizing the work of the maximally loaded site.

In order to decide how to prioritize cost minimization versus load balancing in the model, introduce a scalar $0\leq \lambda \leq 1$ and weight the original cost from \eqref{eq:noloadbalance} and $m$ by $\lambda$ and $(1-\lambda)$, respectively.
The new objective is then
\begin{equation}\label{eq:cost_loadbalance}
\lambda\sum_{a,t,s}c_1(a,t) x_{t,s} y_{a,s} + \lambda\sum_{a,s}c_2(a)  y_{a,s} + (1-\lambda) m
\end{equation}
where $m$ is constrained as follows:
$$
\sum_{a,t} c_3(a,t)x_{t,s} y_{a,s} + \sum_{a,q}c_4(a)y_{a,s} \leq m \quad , \forall s \in \mcS.
$$

Notice that while we are now minimizing \eqref{eq:cost_loadbalance}, the objective of \eqref{eq:noloadbalance} should still be considered as the actual cost of a solution.

\subsection{Linerarization}\label{sec:linearization}
We use the technique discussed in \cite[Chapter IV, Theorem 4]{Hammer68booleanmethods} to linearize the model.
This is done by replacing the quadratic terms in the model with a variable $u_{t,a,s}$ and adding the following new constraints:
\begin{equation*}
  \label{eq:linear}
  \begin{array}{ll}
    u_{t,a,s} \leq  x_{t,s} & \forall t,a,s\\
    u_{t,a,s} \leq  y_{a,s} & \forall t,a,s\\
    u_{t,a,s} \geq  x_{t,s} + y_{a,s} - 1 & \forall t,a,s
  \end{array}
\end{equation*}
For $u_{t,a,s} \geq 0$, notice that $u_{t,a,s} = 1$ if and only if $x_{t,s} = y_{a,s} = 1$ and that $u_{t,a,s}$ is guaranteed to be binary if both $x_{t,s}$ and $y_{a,s}$ are binary (thus, there is no need for requiring it explicitly in the model).

Now, the model in \eqref{eq:noloadbalance} extended with load balancing looks as follows when linearized:
\begin{equation}
\begin{array}{lrll}
\min  & \multicolumn{3}{l}{\lambda\sum_{t,a,s}c_1(a,t) u_{t,a,s} + \lambda\sum_{a,s}c_2(a)  y_{a,s} + (1-\lambda) m}\\
\text{s.t.} & \sum_s x_{t,s} &= 1                       & \forall t\\
            & \sum_s y_{a,s} &\geq 1                         & \forall a\\
            & y_{a,s}-x_{t,s}\varphi_{a,t} &\geq 0            & \forall a, t\\
            & \sum_{a,t} c_3(a,t) u_{a,t,s} + \sum_{a,q}c_4(a)y_{a,s} &\leq m      & \forall s \\
            & u_{t,a,s} - x_{t,s} & \leq 0                    & \forall t, a, s\\
            & u_{t,a,s} - y_{a,s} & \leq 0                    & \forall t, a, s\\
            & u_{t,a,s} - x_{t,s} - y_{a,s} + 1 &\geq  0       & \forall t, a, s\\
            & x_{t,s}, y_{a,s} &\in \{0,1\}                   & \forall t, a, s\\
            & u_{t,a,s} &\geq 0                               & \forall t, a, s
\end{array}
\end{equation}

\subsection{Complexity}\label{sec:complexity}
The objective function in quadratic programs can be written on the form
$$
\tfrac{1}{2}z^\text{T}Qz + cz + d
$$
where in our case $z=(x_{1,1},\dots,x_{|\mcT|,|\mcS|},y_{1,1},\dots,x_{|\mcA|,|\mcS|})$ is a vector containing the decision variables, $Q$ is a cost matrix, $c$ is a cost vector and $d$ a constant.
$Q$ can be easily defined from \eqref{eq:cost_loadbalance} by dividing $Q$ into four quadrants, letting the sub-matrices in the upper-left and lower-right quadrant equal zero and letting the upper-right and lower-left submatrices be defined by $c_1(a,t)$.
$Q$ is indefinite and the cost function \eqref{eq:cost_loadbalance} therefore not convex.
As shown by \citet{qp_nphard} finding optimum when $Q$ is indefinite is NP-hard.

\section{The SA solver -- a heuristic approach}\label{sec:sa}
We develop a heuristic based on simulated annealing (see \cite{wolsey98}) and will refer to it as the \emph{SA}-solver from now on.
The base idea is to alternately fix $x$ and $y$ and only optimize the not-fixed vector, thereby simplifying the problem.
In each iteration we search in the neighborhood of the found solution and accept a worse solution as base for a further search with decreasing probability.

Let $x_{t,s}$ hold an assignment of transactions to sites and define the neighborhood $x'$ of $x$ as a change of location for a subset of the transactions so that for each $t\in \mcT$ we still have $\sum_{s}x'_{t,s} = 1$.
Similarly, let $y_{a,s}$ hold an assignment of attributes to sites but define the neighborhood $y'$ of $y$ as an extended replication of a subset of the attributes.
That is, for each $a\in \mcA$ in that subset we have $y_{a,s}=1 \Rightarrow y'_{a,s}$ and $\sum_{s}y'_{a,s} > \sum_{s}y_{a,s}$.
We found that altering the location for a constant number of 10\% of both transactions/attributes yielded the best results.
The heuristic now looks as pictured in Algorithm \ref{alg:sa}.
\begin{algorithm}
  \caption{The heuristic based on simulated annealing (SA). It iteratively fixes $x$ and $y$ and accepts a worse solution from the neighborhood with decreasing probability.}\label{alg:sa}
\begin{algorithmic}[1]
\State Initialize temperature $\tau>0$ and reduction factor $\rho\in\ ]0;1[$
\State Set the number $L$ of inner loops
\State Initialize $x$ randomly so that \eqref{eq:cons_x} is satisfied
\State \texttt{fix} $\leftarrow$ ``$x$''
\State $S \leftarrow$ findSolution(\texttt{fix})
\While{not frozen}
  \For{$i \in \{1,\dots,L\}$}
    \State $x \leftarrow$ neighborhood of $x$
    \State $y \leftarrow$ neighborhood of $y$
    \State $S' \leftarrow$ findSolution(\texttt{fix})
    \State $\Delta \leftarrow \text{cost}(S') - \text{cost}(S)$
    \State $p \leftarrow$ a randomly chosen number in $[0;1]$
    \If{$\Delta \leq 0$ or $p < e^{-\Delta/\tau}$}
      \State $S\leftarrow S'$
    \EndIf
    \State \texttt{fix} $\leftarrow$ the element in $\{$``$x$''$,$``$y$''$\} \setminus \{\texttt{fix}\}$
  \EndFor
  \State $\tau \leftarrow \rho\cdot \tau$
\EndWhile
\end{algorithmic}
\end{algorithm}
Notice, that the linearization constraints is not needed since either $x$ or $y$ will be constant in each iteration.
This reduces the size of the problem considerably.


\section{Further improvements}\label{sec:improvements}
Consider a table with $n$ attributes together with two queries: one accessing attribute 1 through $k$ and one accessing attribute $k$ through $n$.
Then it is sufficient to find an optimal distribution for the three attribute groupings $\{1,\dots,k-1\}$, $\{k\}$ and $\{k+1,\dots,n\}$, considering each group as an atomic unit and thereby reducing the problem size.
In general, it is only necessary to distribute groups of attributes induced by query access overlaps.
\citet{chu93} refer to these attribute overlaps as \emph{reasonable cuts}.
Even though this will not improve the worst-case complexity, this reduction may still have a large performance impact on some instances.

Also, assuming that transactions follow the 20/80 rule (20\% of the transactions generate 80\% of the load), the problem can be solved iteratively over $\mcT$ starting with a small set of the most heavy transactions.

\section{Computational results}\label{sec:results}
We assume that the context is a database with a very high transaction count like the memory-only database H-store \cite{hstore} (now VoltDB\footnote{http://voltdb.com}) and thus need to compare RAM access versus network transfer time when deciding an appropriate network penalty factor $p$.
A PCI Express 2.0 bus transfers between 32 Gbit/s and 128 Gbit/s while the bandwidth of PC3 DDR3-SDRAM is at least 136 Gbit/s so the bus is the bottleneck in RAM accesses.
We assume that the network is well configured and latency is minimal.
Therefore the network penalty factor could be estimated to $p\in [3;128]$ if either a gigabit or 10-gigabit network is used to connect the physical sites.
We assume the use of a 10-gigabit network and therefore set $p=8$ in our tests unless otherwise stated.

We furthermore mainly focus on minimizing the total costs of execution and therefore set $\lambda$ low.
If $\lambda$ is kept positive the model will, however, choose the more load balanced layout if there is a cost draw between multiple layouts.
We set $\lambda = 0.1$ in our tests unless otherwise stated.

All tests were run on a MacBook Pro with a 2.4 GHz Intel Core 2 Duo and 4GB 1067 MhZ DDR3 RAM, running Mac OS X 10.5.
The GNU Linear Programming Kit\footnote{http://gnu.org/software/glpk} (GLPK) 4.39 was used as MIP solver, using only a single thread.

The test implementation is available upon request.

\subsection{Initial temperature}
The temperature $\tau$ used in the heuristic described in Section \ref{sec:sa} determines how willing the algorithm is to accept a worse solution than the currently best found.
Let $C^*$ and $C$ denote the objective for the best solution so far and the currently generated solution, respectively.
In the computational results provided here we accept a worse solution with 50\% probability in the first set of iterations if $\tfrac{C-C^*}{C} < 5\%$.
Referring to the notation used in Algorithm \ref{alg:sa}, we have $50\% = e^{0.05C^* / \tau}$ and thus an initial temperature of $\tau = -0.05C^* / \ln 0.5$.

\subsection{The TPC-C v5 instance}
We perform tests on the TPC-C version 5.10.1 benchmark\footnote{http://www.tpc.org/tpcc}.
The TPC-C specification describes transactions, queries and database schema but does not provide the statistics needed to create a problem instance.
We therefore made some simplified assumptions:
all queries are assumed to run with equal frequency and all queries (not transactions) are assumed to access a single row except in the obvious cases where aggregates are used or there are being iterated over the result.
In these cases we assume that the query accesses 10 rows.
Thereby, the New-Order transaction for example, are assumed to access 11 rows in average.

We model UPDATE queries as two sub-queries: A read-query accessing all the attributes used in the original query and a write-query only accessing the attributes actually being written (and thus whose update needs to be distributed to all replicas).

\subsection{Random instances}
To the best of our knowledge there is no standard library of typical OLTP instances with schemas, workloads and statistics so in order to explore the characteristics of the algorithms we perform some experiments on a set of randomly generated instances instead as it showed up to be a considerable administrative and bureaucratic challenge (if possible at all) to collect appropriate instances from ``real life'' databases.
The randomly generated instances vary in several parameters in order to clarify which characteristics that influence the potential cost reduction by applying our vertical partitioning algorithms.
The parameters include: number of transactions in workload, number of tables in schema, maximum number of attributes per table, maximum number of queries per transaction, percentage of queries being updates, maximum number of different tables being referred to from a single query, maximum number of individual attributes being referred to by a single query, the set of allowed attribute widths.
We define classes of problem instances by upper bounds on all parameters.
Individual instances are then generated by choosing the value of each parameter evenly distributed between 1 and its upper bound.
That is, if e.g. the maximum allowed number of attributes in tables is $k$, the number of table attributes for each table in the generated instance will be evenly distributed between 1 and $k$ with a mean of $k/2$.

\subsection{Results}
In the following we perform a series of tests and display the results in tables where each entry holds the found objective of \eqref{eq:noloadbalance} for the given instance.

Table \ref{fig:results_sa_params} explores the influence of a set of parameters in the randomly generated instances by varying one parameter at a time while fixing the rest.
We test two classes of instances using the SA solver: a smaller with \#tables = $|\mcT|$ = 20 and a larger with \#tables = $|\mcT|$ = 100.
The results suggest that the largest workload reduction is obtained for instances having relatively few queries per transaction, few updates, many attributes per table and/or a moderate number of attribute references per query.
The number of table references per query and the allowed attribute widths, however, only seem to have moderate influence on the result.
\newlength{\colwidth}
\setlength{\colwidth}{2.5cm}
\begin{table}[htp]
  \centering
  \footnotesize
\begin{tabular}{lr|ccc|ccc}
   &     & \multicolumn{3}{c|}{\#tables = $|\mcT|$ = 20} & \multicolumn{3}{c}{\#tables = $|\mcT|$ = 100}\\
   &     & $|\mcS| = 1$ & $|\mcS| = 2$ & $|\mcS| = 3$ & $|\mcS| = 1$ & $|\mcS| = 2$ & $|\mcS| = 3$\\
\hline
\multirow{3}{\colwidth}{\textbf{A}\quad Max queries per transaction} &           1 & 0.585 & 0.309 & 0.278    &   3.194 & 1.784 & 1.471 \\
                                                     &  \textbf{3} & 1.567 & 1.478 & 1.386    &   5.743 & 4.550 & 4.189 \\
                                                     &           5 & 1.305 & 1.054 & 0.972    &   8.840 & 7.569 & 6.983 \\
\hline
\multirow{3}{\colwidth}{\textbf{B}\quad Percent updates queries} &           0 & 1.747 & 1.369 & 1.110    &   5.959 & 4.235 & 3.510 \\
                                                     & \textbf{10} & 1.567 & 1.478 & 1.386    &   5.743 & 4.550 & 4.189 \\
                                                     &          30 & 1.349 & 1.244 & 1.263*   &   5.106 & 4.555 & 4.462\\
\hline
\multirow{3}{\colwidth}{\textbf{C}\quad Max attributes per table}   &           5 & 0.520 & 0.520*& 0.520*   &   2.583 & 2.772*& 2.712*\\
                                                     & \textbf{15} & 1.567 & 1.478 & 1.386    &   5.743 & 4.550 & 4.189 \\
                                                     &          35 & 1.643 & 0.968 & 0.850    &  14.970 & 7.341 & 5.355\\
\hline
\multirow{3}{\colwidth}{\textbf{D}\quad Max table re\-fe\-ren\-ces per query} &    2 & 0.602 & 0.430 & 0.356    &   3.447 & 3.022 & 2.865 \\
                                                     &  \textbf{5} & 1.567 & 1.478 & 1.386    &   5.743 & 4.550 & 4.189 \\
                                                     &          10 & 2.246 & 1.607 & 1.516    &   8.147 & 6.063 & 5.623 \\
\hline
\multirow{3}{\colwidth}{\textbf{E}\quad Max attribute references per query} &   5 & 0.678 & 0.288 & 0.199    &   5.176 & 2.526 & 1.969 \\
                                                     & \textbf{15} & 1.567 & 1.478 & 1.386    &   5.743 & 4.550 & 4.189 \\
                                                     &          25 & 1.115 & 0.988 & 1.008*   &   5.641 & 5.909*& 5.684* \\
\hline
\multirow{3}{\colwidth}{\textbf{F}\quad Allowed attribute widths}   & $\{2,4,8\}$ & 1.194 & 1.080 & 1.030    &   4.456 & 3.488 & 3.500*\\
                                              & $\mathbf{\{4,8\}}$ & 1.567 & 1.478 & 1.386    &   5.743 & 4.550 & 4.189 \\
                                                    & $\{4,8,16\}$ & 2.387 & 2.160 & 2.060    &   8.912 & 6.977 & 7.000\\
\end{tabular}
\caption{Comparing the effect of parameter changes. Results were found using the SA solver. We test three possible values for each parameter, varying one parameter at the time and fixing all other parameters at their default value (marked with bold). The costs are shown in units of $10^6$. Tests are divided into two classes having both the number of transactions and schema tables equal to 20 (left) and 100 (right), respectively.
The results suggest that the largest workload reduction, unsurprisingly, is obtained for instances having relatively few queries per transaction, few updates, many attributes per table and/or a moderate number of attribute references per query. The number of table references per query and the allowed attribute widths, however, only seem to have moderate influence on the result.
} \label{fig:results_sa_params}
\end{table}

Table \ref{fig:results_qp_sa} compares the QP and SA solvers on the TPC-C benchmark and a set of randomly generated larger instances, divided into two classes with either large or low potential for cost reduction.
The random instances are described in Table \ref{fig:results_rnd_instances} where the columns here refer to the single-letter labels for the parameters shown in Table \ref{fig:results_sa_params}.
\begin{table}[htp]
  \centering
  \footnotesize
\begin{tabular}{l|cccccccc}
Name        & A & B  & C  & D & E & F              & $|\mcT|$ & \#tables\\
\hline
rndAt4x15    & 3 & 10 & 30 & 3 & 8 & $\{2,4,8,16\}$ & 15    & 4\\
rndAt8x15    & 3 & 10 & 30 & 3 & 8 & $\{2,4,8,16\}$ & 15    & 8\\
rndAt8x15u50 & 3 & 50 & 30 & 3 & 8 & $\{2,4,8,16\}$ & 15    & 8\\
rndAt16x15   & 3 & 10 & 30 & 3 & 8 & $\{2,4,8,16\}$ & 15    & 16\\
rndAt32x15   & 3 & 10 & 30 & 3 & 8 & $\{2,4,8,16\}$ & 15    & 32\\
\hline
rndAt4x100      & 3 & 10 & 30 & 3 & 8 & $\{2,4,8,16\}$ & 100    & 4\\
rndAt8x100      & 3 & 10 & 30 & 3 & 8 & $\{2,4,8,16\}$ & 100    & 8\\
rndAt16x100     & 3 & 10 & 30 & 3 & 8 & $\{2,4,8,16\}$ & 100    & 16\\
rndAt32x100     & 3 & 10 & 30 & 3 & 8 & $\{2,4,8,16\}$ & 100    & 32\\
\hline
rndBt4x15      & 3 & 10 &  5 & 6 & 28 & $\{2,4,8,16\}$ & 15    & 4\\
rndBt8x15      & 3 & 10 &  5 & 6 & 28 & $\{2,4,8,16\}$ & 15    & 8\\
rndBt16x15     & 3 & 10 &  5 & 6 & 28 & $\{2,4,8,16\}$ & 15    & 16\\
rndBt16x15u50  & 3 & 50 &  5 & 6 & 28 & $\{2,4,8,16\}$ & 15    & 16\\
rndBt32x15     & 3 & 10 &  5 & 6 & 28 & $\{2,4,8,16\}$ & 15    & 32\\
\hline
rndBt4x100      & 3 & 10 &  5 & 6 & 28 & $\{2,4,8,16\}$ & 100   & 4\\
rndBt8x100      & 3 & 10 &  5 & 6 & 28 & $\{2,4,8,16\}$ & 100   & 8\\
rndBt16x100     & 3 & 10 &  5 & 6 & 28 & $\{2,4,8,16\}$ & 100   & 16\\
rndBt32x100     & 3 & 10 &  5 & 6 & 28 & $\{2,4,8,16\}$ & 100   & 32
\end{tabular}
\caption{Random instances used when comparing the QP and SA solvers in Table \ref{fig:results_qp_sa}. The instances in the upper part (rndA\dots) are expected to get a large cost reduction while instances in the lower part (rndB\dots) are expected to get a small cost reduction. The columns refer to the single-letter labels for the parameters shown in Table \ref{fig:results_sa_params}.}\label{fig:results_rnd_instances}
\end{table}
\begin{table}[htp]
  \centering
  \footnotesize
\begin{tabular}{l|ccc|cc|cc|c}
              &          &       &          & \multicolumn{2}{c|}{QP} & \multicolumn{2}{c|}{SA} &\\
  Instance    & $|\mcA|$ & $|\mcT|$ & $|\mcS|$ & Cost    & Time (s) & Cost   & Time (s) & $|\mcS|=1$ \\
\hline
  TPC-C v5    & 92       & 5     & 2       & \textbf{0.133}  &    1 & 0.138 &    5 & 0.208\\ %
  TPC-C v5    & 92       & 5     & 3       & \textbf{0.132}  &    6 & \textbf{0.132} &    5 & 0.208\\ %
  TPC-C v5    & 92       & 5     & 4       & \textbf{0.132}  &   33 & \textbf{0.132} &    5 & 0.208\\ %
\hline\hline
  rndAt4x15      & 54       & 15    & 4       &\textbf{(0.332)} &  1800 & 0.396 &   10  & 0.933   \\ %
  rndAt8x15      & 105      & 15    & 4       &\textbf{(0.324)} &  1800 & 0.327 &   18  & 0.808 \\ %
  rndAt16x15     & 225      & 15    & 4       &\textbf{(0.267)} &  1800 & 0.309 &   41  & 1.180 \\ %
  rndAt32x15     & 492      & 15    & 4       &(0.315) &  1800 & \textbf{0.217} &  89   & 1.491 \\ %
  rndAt64x15     & 1023     & 15    & 4       &\textbf{(0.269)} &  1800 & 0.268 &  190& 1.452\\
\hline
  rndAt4x100     & 54       & 100   & 4       &(8.001) &1800  &  8.246 &  79 & \textbf{7.946} \\ 
  rndAt8x100     & 105      & 100   & 4       &(7.681) &1800  &  8.018 & 150 & \textbf{7.454} \\ 
  rndAt16x100    & 225      & 100   & 4       &-       &t/o   &  \textbf{6.525} & 321 & 8.741 \\ %
  rndAt32x100    & 492      & 100   & 4       &-       &t/o   &  \textbf{4.501} & 728 & 8.916 \\ %
  rndAt64x100    & 1023     & 100   & 4       &-       &t/o   &  \textbf{4.119} & 1531& 9.591 \\
\hline\hline
  rndBt4x15      & 12      & 15    & 4        &  0.303 &   65 & 0.303 & 3    & \textbf{0.303} \\ %
  rndBt8x15      & 27      & 15    & 4        &(0.448) & 1800 & \textbf{0.424} & 6    & 0.440\\ %
  rndBt16x15     & 49      & 15    & 4        &\textbf{(0.333)} & 1800 & 0.334 & 9    & 0.385 \\ %
  rndBt32x15     & 98      & 15    & 4        &\textbf{(0.319)} & 1800 & \textbf{0.319} & 16   & 0.361 \\ %
  rndBt64x15     & 210     & 15    & 4        &\textbf{(0.221)} & 1800 & \textbf{0.221} & 31   & 0.229 \\
\hline
  rndBt4x100     & 54       & 100   & 4       &(4.484) &1800  & \textbf{2.251}  & 18    & \textbf{2.251} \\ %
  rndBt8x100     & 105      & 100   & 4       &(4.323) &1800  & \textbf{2.419}  & 37    &  \textbf{2.419} \\ %
  rndBt16x100    & 225      & 100   & 4       &(2.001) &1800  & \textbf{1.774}  & 62    & \textbf{1.774} \\ %
  rndBt32x100    & 492      & 100   & 4       &(2.419) &1800  & \textbf{1.999}  & 124   & \textbf{1.999} \\ %
  rndBt64x100    & 1023     & 100   & 4       &-       &1800  & \textbf{2.473}  & 270   & \textbf{2.473} \\
\end{tabular}
\caption{Comparing the QP algorithm with the simulated annealing based heuristic (SA), allowing attribute replication and with remote partition placement. Costs are shown in units of $10^6$. The SA algorithm had a 30 second time limit for each iteration and if the limit was reached it proceeded with another neighborhood. The QP algorithm had a time bound of 30 minutes and an MIP tolerance gap of 0.1\%. Where the time limit was reached, the best found cost (if any) is written in parentheses. ``t/o'' indicates that no integer solution was found within the time limit.} \label{fig:results_qp_sa}
\end{table}
As seen in Table \ref{fig:results_qp_sa} the SA solver is generally faster than the QP solver but the QP solver obtains lower costs when the instances are small.
Expectedly, the instances in class ``rndB\dots'' with many attribute references per query but few queries per table gains little or no cost reduction by applying the algorithms.
TPC-C, on the other hand, gets a cost reduction of 37\% and the random instances in class ``rndA\dots'', with many attributes per table and relatively few attribute references per query, get a cost reduction between 25\% and 85\%.
None of the algorithms found a cost reduction for the instances rndAt4x100 and rndAt8x100 because of the ``overweight'' of transactions compared to the number of attributes in the schemas.

Table \ref{fig:tpcc-result} depicts an actual partitioning of TPC-C constructed by the QP solver for three sites.
\begin{table}[htp]
  \centering
  \tiny
\begin{minipage}[t]{.327\textwidth}
\vspace{0pt}
\begin{tabular}{|l|}
\hline
{\bf Site 1}\\
\hline
\hline
Transaction \emph{Payment}\\
\hline
\hline
Customer.C\_BALANCE\\
Customer.C\_CITY\\
Customer.C\_CREDIT\\
Customer.C\_CREDIT\_LIM\\
Customer.C\_DATA\\
Customer.C\_DISCOUNT\\
Customer.C\_D\_ID\\
Customer.C\_FIRST\\
Customer.C\_ID\\
Customer.C\_LAST\\
Customer.C\_MIDDLE\\
Customer.C\_PHONE\\
Customer.C\_SINCE\\
Customer.C\_STATE\\
Customer.C\_STREET\_1\\
Customer.C\_STREET\_2\\
Customer.C\_W\_ID\\
Customer.C\_ZIP\\
\hline
District.D\_CITY\\
District.D\_ID\\
District.D\_NAME\\
District.D\_STATE\\
District.D\_STREET\_1\\
District.D\_STREET\_2\\
District.D\_W\_ID\\
District.D\_YTD\\
District.D\_ZIP\\
\hline
History.H\_AMOUNT\\
History.H\_C\_D\_ID\\
History.H\_C\_ID\\
History.H\_C\_W\_ID\\
History.H\_DATA\\
History.H\_DATE\\
History.H\_D\_ID\\
History.H\_W\_ID\\
\hline
OrderLine.OL\_DIST\_INFO\\
OrderLine.OL\_NUMBER\\
\hline
Stock.S\_ORDER\_CNT\\
Stock.S\_REMOTE\_CNT\\
Stock.S\_YTD\\
\hline
Warehouse.W\_CITY\\
Warehouse.W\_ID\\
Warehouse.W\_NAME\\
Warehouse.W\_STREET\_1\\
Warehouse.W\_STREET\_2\\
Warehouse.W\_YTD\\
Warehouse.W\_ZIP\\
\hline
\end{tabular}
\end{minipage}
\hfill
\begin{minipage}[t]{.327\textwidth}
\vspace{0pt}
\begin{tabular}{|l|}
\hline
{\bf Site 2}\\
\hline
\hline
Transaction \emph{StockLevel}\\
\hline
\hline
Customer.C\_CITY\\
Customer.C\_DELIVERY\_CNT\\
Customer.C\_PAYMENT\_CNT\\
Customer.C\_SINCE\\
Customer.C\_YTD\_PAYMENT\\
\hline
District.D\_ID\\
District.D\_NEXT\_O\_ID\\
District.D\_W\_ID\\
\hline
Item.I\_IM\_ID\\
\hline
OrderLine.OL\_D\_ID\\
OrderLine.OL\_I\_ID\\
OrderLine.OL\_O\_ID\\
OrderLine.OL\_W\_ID\\
\hline
Stock.S\_I\_ID\\
Stock.S\_QUANTITY\\
Stock.S\_W\_ID\\
\hline
\end{tabular}
\end{minipage}
\hfill
\begin{minipage}[t]{.327\textwidth}
\vspace{0pt}
\begin{tabular}{|l|}
\hline
{\bf Site 3}\\
\hline
\hline
Transaction \emph{Delivery}\\
Transaction \emph{NewOrder}\\
Transaction \emph{OrderStatus}\\
\hline
\hline
Customer.C\_BALANCE\\
Customer.C\_CREDIT\\
Customer.C\_DISCOUNT\\
Customer.C\_D\_ID\\
Customer.C\_FIRST\\
Customer.C\_ID\\
Customer.C\_LAST\\
Customer.C\_MIDDLE\\
Customer.C\_W\_ID\\
\hline
District.D\_ID\\
District.D\_NEXT\_O\_ID\\
District.D\_TAX\\
District.D\_W\_ID\\
\hline
Item.I\_DATA\\
Item.I\_ID\\
Item.I\_NAME\\
Item.I\_PRICE\\
\hline
NewOrder.NO\_D\_ID\\
NewOrder.NO\_O\_ID\\
NewOrder.NO\_W\_ID\\
\hline
Order.O\_ALL\_LOCAL\\
Order.O\_CARRIER\_ID\\
Order.O\_C\_ID\\
Order.O\_D\_ID\\
Order.O\_ENTRY\_D\\
Order.O\_ID\\
Order.O\_OL\_CNT\\
Order.O\_W\_ID\\
\hline
OrderLine.OL\_AMOUNT\\
OrderLine.OL\_DELIVERY\_D\\
OrderLine.OL\_D\_ID\\
OrderLine.OL\_I\_ID\\
OrderLine.OL\_O\_ID\\
OrderLine.OL\_QUANTITY\\
OrderLine.OL\_SUPPLY\_W\_ID\\
OrderLine.OL\_W\_ID\\
\hline
Stock.S\_DATA\\
Stock.S\_DIST\_01\\
Stock.S\_DIST\_02\\
Stock.S\_DIST\_03\\
Stock.S\_DIST\_04\\
Stock.S\_DIST\_05\\
Stock.S\_DIST\_06\\
Stock.S\_DIST\_07\\
Stock.S\_DIST\_08\\
Stock.S\_DIST\_09\\
Stock.S\_DIST\_10\\
Stock.S\_I\_ID\\
Stock.S\_QUANTITY\\
Stock.S\_W\_ID\\
\hline
Warehouse.W\_ID\\
Warehouse.W\_TAX\\
\hline
\end{tabular}
\end{minipage}
\caption{The result of a vertical partitioning of the TPC-C benchmark using the QP solver for three sites.
Each column represents the contents of a site and is divided into three sub-sections: a header, a section holding the transaction names and a longer section holding the attributes assigned to the respective site.} \label{fig:tpcc-result}
\end{table}

Table \ref{fig:results_replication} illustrates the effect of disjoint versus nondisjoint partitioning, that is, partitioning without and with attribute replication.
As seen, greater cost reduction can be obtained when allowing replication but in exchange to increased computation time.
\begin{table}[htp]
  \centering
  \footnotesize
\begin{tabular}{l|ccc|cc|cc|c}
              &          &       &          & \multicolumn{2}{c|}{w. replication} & \multicolumn{2}{c|}{w/o replication} & \\
  Instance    & $|\mcA|$ & $|\mcT|$ & $|\mcS|$ & Cost    & Time (s) & Cost   & Time (s) & Ratio\\
\hline
  TPC-C v5    & 92       & 5     & 1        & 0.208  &    0 & 0.208 &    0 &        - \\
  TPC-C v5    & 92       & 5     & 2        & 0.133  &    1 & 0.207 &    1 &      64\% \\
  TPC-C v5    & 92       & 5     & 3        & 0.132  &    6 & 0.207 &    2 &      64\% \\
  TPC-C v5    & 92       & 5     & 4        & 0.132  &   33 & 0.207 &    3 &      64\% \\
\hline
  rndAt4x15   & 54       & 15    & 2        & 4.855  &   28 & 6.799 &    1 &     71\% \\
  rndAt8x15   & 105      & 15    & 2        & 4.710  &  517 & 5.809 &    6 &     81\% \\
  rndAt8x15   & 27       & 15    & 2        & 4.244  &   4  & 4.402 &    0 &     96\% \\
  rndAt16x15  & 49       & 15    & 2        & 3.410  &  34  & 3.852 &    0 &     89\% \\
\end{tabular}
\caption{Computational results from solving the TPC-C benchmark and a few random instances with the QP solver. Costs are shown in units of $10^5$. The table shows that costs can be reduced by allowing attribute replication and that TPC-C does not benefit noticeably from being partitioned and distributed to more than two sites.  The \emph{Ratio} column displays the ratio between the replicated and non-replicated cost.} \label{fig:results_replication}
\end{table}

Table \ref{fig:results_placement} compares two different kinds of partition placements: 1) all partitions being located at one single site (thereby avoiding inter-site transfers) and 2) partitions being located at remote sites.
These two situations can be simulated by setting $p=0$ and $p>0$, respectively.
The benefits of local placements are given by the amount of updates in the workload as only updates cause inter-site transfers.
More updates implies larger costs for remote placements.
For a somewhat extreme case, instance ``rndAt8x15u50'', with 50\% of the queries being updates, the costs are about 33\% lower when placing the partitions locally.
\begin{table}[htp]
  \centering
  \footnotesize
\begin{tabular}{l|ccc|cc|cc}
              &          &       &          & \multicolumn{2}{c|}{Local} & \multicolumn{2}{c}{Remote} \\
  Instance    & $|\mcA|$ & $|\mcT|$ & $|\mcS|$ & Cost (QP)  & Cost (SA) & Cost (QP) & Cost (SA)\\
\hline
  TPC-C v5    & 92       & 5     & 1        & 1.916    & 1.916   & 1.916  & 1.916 \\
  TPC-C v5    & 92       & 5     & 2        & 1.210    & 1.208   & 1.221  & 1.273 \\
  TPC-C v5    & 92       & 5     & 3        & 1.208    & 1.208   & 1.220  & 1.220 \\
\hline
  rndAt4x15     & 54       & 15    & 2        & 4.709    & 4.742   & 4.855  & 4.888 \\
  rndAt8x15     & 105      & 15    & 2        & 4.424    & 4.808   & 4.710  & 5.187 \\
  rndAt8x15u50  & 105      & 20    & 2        & 3.189    & 3.313   & 4.778  & 4.873 \\
  rndBt8x15     & 27       & 15    & 2        & 4.365    & 4.332   & 4.244  & 4.730 \\ 
  rndBt16x15    & 49       & 15    & 2        & 3.335    & 3.387   & 3.410  & 3.404\\
  rndBt16x15u50 & 49       & 20    & 2        & 5.066    & 5.220   & 5.438  & 5.438
\end{tabular}
\caption{Comparing the costs of local ($p=0$) versus remote ($p>0$) location of partitions and with attribute replication allowed. Costs are in units of $10^5$. Write-rarely instances or instances in class ``rndB\dots'' do not benefit noticeably by placing all partitions locally, even the instances with 50\% update queries, however instances in class ``rndA\dots'' with a large update ratio do. The reason is that only updates cause inter-site transfer. That the costs of the local placement for rndBt8x15 is \emph{larger} than when placed remotely is since $\lambda>0$.} \label{fig:results_placement}
\end{table}

\section{Conclusion}\label{sec:conclusion}
We have constructed a cost model for vertical partitioning of relational OLTP databases together with a quadratic integer program that distributes both attributes and transactions to a set of sites while allowing attribute replication, preserving single-sitedness for read queries and in which load balancing vs. total cost minimization can be prioritized arbitrarily.

We also presented a more scalable heuristic which seems to deliver good results.
For both algorithms we obtained a cost reduction of 37\% in our model of TPC-C and promising results for the random instances.
Even though the latter theoretically can be constructed with arbitrary high/low benefits from vertical partitioning, the test runs on our selected subset of random instances seem to indicate that 1)~our heuristic scales far better than the QP-solver, and 2)~it can obtain valuable cost reductions on many real-world OLTP databases, as we tried to select the parameters realistically.

One thing we miss, however, is an official OLTP testbed -- a library containing realistic OLTP workloads, schemas and statistics.
Such a collection of realistic instances could serve as base for several insteresting and important studies for understanding the nature and characteristics of OLTP databases.

\section*{Acknowledgements}
The author would like to acknowledge Daniel Abadi for competent and valuable discussions and feedback.
Also, Rasmus Pagh, Philippe Bonnet and Laurent Flindt Muller have been very helpful with insightful comments on preliminary versions of the paper.

\clearpage
\bibliographystyle{plainnat}
\bibliography{bibliography}
\clearpage
\appendix
\section{Latency}\label{sec:latency}
This section describes how to extend the algorithms to also estimate costs of network latency for queries accessing attributes on remote sites.
We assume, that all remote access (if any) for queries are done in parallel and with a constant number of requests per query per remote site.
Let $p_l$ denote a latency penalty factor and introduce a new binary variable $\psi_q$ for each query $q$ indicating with $\psi_q=1$ if $q$ accesses any remotely placed attributes.
Letting $n$ denote the number of remotely accessed attributes by $q$ we have $n>0\Rightarrow \psi_q =1$ and $n=0\Rightarrow \psi_q = 0$, or equivalently $(\psi_q-1)n = 0$ and $\psi_q - n \leq 0$.
This results in the following two classes of new constraints:
$$
(\psi_q-1)\sum_{a,s} \delta_q \alpha_{a,q} \gamma_{q,t}(1-x_{t,s}) y_{a,s} = 0\quad, \forall q,t
$$
and
$$
\psi_q - \sum_{a,s} \delta_q \alpha_{a,q} \gamma_{q,t}(1-x_{t,s}) y_{a,s} \leq 0\quad, \forall q,t
$$
The total latency in a given partitioning can now be estimated by the sum $p_l\sum_{q}f_q\psi_q$ which can be added to the cost objective function \eqref{eq:noloadbalance}.

\end{document}